\documentclass[12pt, aps, showpacs]{revtex4}
\usepackage{hhline}
\begin{document}
\title{Effects of core polarization on the nuclear Schiff moment}
\author{ V.F. Dmitriev, R.A. Sen'kov }
\affiliation{Budker Institute of Nuclear Physics, Novosibirsk-90, 630090 Russia}
\affiliation{Novosibirsk State University, Novosibirsk-90, 630090 Russia } 
\author{N. Auerbach}
\affiliation{School of Physics and Astronomy, Tel Aviv University,
 Tel-Aviv,  69978 Israel}
\begin{abstract}
Schiff moments were calculated for a set of nuclei with full account of core polarization effects. A finite range P and T violating weak nucleon-nucleon interaction has been used in the calculations.
While in the absence of core polarization the Schiff moment depends on one combination of the weak interaction constants, in the presence of core polarization the Schiff moment depends on all three constants separately. The dominant contribution comes from isovector,  $\Delta T=1$, part of the weak interaction. The effects of core polarization were found to have in general a large effect on the Schiff moments.
\end{abstract}
\pacs{21.10.Ky, 21.60.Iz}
\maketitle
\section{Introduction}
The theory  of static atomic dipole moments relies mainly on presence of parity (P) and time (T) reversal violating electro-magnetic (E-M) moments generated by nuclei. It is conjectured that the dominant contribution to the atomic dipole moment comes from parity and time reversal violating components in the nucleon-nucleon (N-N) force. These  components of the interaction induce P-T violating E-M  moments in the nucleus. In turn these forbidden nuclear moments produce parity and time reversal violation in the E-M  field  which  can induce  static electric dipole moments in the atom. This way of reasoning is the  working  assumption in the study of the atomic dipole moment \cite{Jac95,Fla86}.  

The measurements of the atomic dipole moment are presently very advanced and reached a very high degree of precision \cite{Jac95, rgf01}.  These experiments are performed with neutral atoms. It has been shown years ago \cite{sch63} that in such cases the field of the electric dipole moment of the nucleus is screened by the field of the electrons and there is no effect when the atom is placed in an external electric field. This situation applies strictly only when one considers point-like particles. When the finite size of the nucleus is taken into account one can  go to higher orders in the expansion of the E-M  field  and find that in the next order in the expansion it is the Schiff moment \cite{fks84} that can induce an electric dipole moment in the atom. (See next section). 

  The upper limits for the existence of dipole moments obtained in the atomic measurements cannot provide directly limits on the presence of  T and P violation in the nuclear Hamiltonian. One needs theoretical input. In order to make the connection one must have precise calculations of moments in nuclei  produced by the T-P non-conserving part of the N-N interaction.  

In the past the  theory of Schiff moments in spherical nuclei was limited  to either a single-particle value (in Z-odd, N-even nuclei) or  calculated for a simple two particle -- one hole (2p-1h)  configuration  (in Z-even, N-odd  nuclei).

In this paper we will go beyond these simple calculations and include various 
effects of core polarization. As we will see these effects are large compared to the
results obtained in lowest order.

We should stress that an important aspect of this work is the use of a finite range interaction for the P-T-violating component of the N-N force.
A finite range interaction was used recently in the calculation of the Schiff moment   of $^{199}$Hg \cite{ds1} and $^{225}$Ra\cite{eng03}. In the present work we examine the influence of finite range for a number of nuclei and  discuss  more  general  features that this leads to. 
   
 We  compute the Schiff moments for the following nuclei, for reasons explained below.
\subsection{Neutron-odd nuclei}
\paragraph{$^{199}$Hg:}  The most accurate upper limit  on parity and time reversal violation has been obtained from the measurement of the atomic electric dipole moment in the Hg atom  \cite{rgf01}.  The nucleus has (Z=80) has an odd number of neutrons. The lowest order contribution to the Schiff moment comes from a 2p-1h configuration (a proton 1p-1h added to the neutron ) admixed into the ground state  (g.s.). Because of the importance of this nucleus in the present experimental studies we have examined in more detail the influence of core polarization effects  on the Schiff moment .
\paragraph{$^{129}$Xe:} The A=129 isotope of  Xenon is also an even-Z,  odd-N nucleus and the  lowest order contribution to the Schiff moment comes from a 1p-1h proton excitation of the core. Similarly to $^{199}$Hg the atom of $^{129}$Xe is used in experimental studies of the atomic electric dipole moment \cite{Jac95}. The upper limit for the dipole moment obtained in the experiment is only several times higher than in Hg. We have, therefore as in the case of Hg, found it useful to  calculate more  accurately the Schiff moment by introducing several effects coming from core polarization and not included in previous theoretical studies.
\paragraph{$^{211}$Rn {\rm and} $^{213}$Ra:}
The light isotopes of radon and radium are considered to be transitional nuclei, still  being close to spherical in their ground states \cite{moll95,moll97}. The heavier isotopes of these elements (A=221-225) are found to have quadrupole  plus octupole deformed shapes in the ground states \cite{moll95,moll97}. It was suggested that in these heavier isotopes  of Rn and Ra  (as well as some other elements in this region ) the Schiff moments  are strongly enhanced because of the quadrupole+octupole deformations \cite{auer66,spe97}. It is therefore useful for the purpose of comparison to calculate accurately the Schiff moments in the neighboring, lighter isotopes where the onset of deformations has not yet occurred.
\paragraph{$^{225}$Ra:}
Here we calculate the Schiff moment assuming that the nucleus is spherical. We do this in order to have a reference value so that one can see the degree of enhancement when a full calculation , that includes deformations, is performed.
\subsection{Proton-odd nuclei}
In a Z-odd, N-even nucleus the odd proton can contribute directly to the Schiff moment. It is therefore usually assumed that the dominant contribution is the single-particle  value of the last proton.
\paragraph{$^{133}$Cs:}
We examined this point for $^{133}$Cs (Z=55) where the core-polarization correction was evaluated together with the single-particle Schiff moment.
\paragraph{$^{223}$Fr:}
This nucleus (Z=87) was found to have an enhanced Schiff moment  when the quadrupole+octupole  deformations were taken into account \cite{spe97}. As  in the case of $^{225}$Ra for the same reason we computed the single-particle and core polarization values taking $^{223}$Fr to be spherical.
\section{P and T violating nucleon-nucleon interaction}
The expectation value of the Schiff moment operator \cite{fks84}
\begin{equation} \label{00}
\hat{\bf S}=\frac{1}{10}\sum_ie_i(r_i^2{\bf r}_i-\frac{5}{3}<r^2>_{ch}{\bf
r}_i) 
\end{equation}
can be non zero only in the presence of P and T violating nuclear forces.
There are good reasons to assume that the exchange of a
$\pi^0$-meson is the most efficient mechanism of generating CP odd
nuclear forces. This is due to the large value of the strong $\pi
NN$ coupling constant $g_s = 13.5$ and to the small pion mass, as
well as to the fact that outer proton and neutron orbitals in
heavy nuclei are quite different. The P odd, T odd effective $\pi
NN$ Lagrangians are conveniently classified by their isotopic
properties \cite{bar,hahe,herc}:
\begin{equation}\label{0}
\Delta T =0. \quad L_0\,=\,g_0\,[\,\sqrt{2}\,(\overline{p}n
\,\pi^+ +\overline{n}p\, \pi^-)\,+(\,\overline{p}p
-\overline{n}n)\,\pi^0];
\end{equation}
\begin{equation}\label{1}
|\Delta T| = 1. \quad L_1\,=\,g_1\,\overline{N}N \,\pi^0 =
\,g_1\,(\,\overline{p} p \,+ \overline{n}n )\,\pi^0;
\end{equation}
\begin{equation}\label{2}
|\Delta T| =2. \quad L_2\,=\,g_2\,[\,\sqrt{2}\,(\overline{p}n
\,\pi^+ +\overline{n}p\, \pi^-)\,- 2(\,\overline{p}p
-\overline{n}n)\,\pi^0].
\end{equation}
The above Lagrangians generate a P and T odd nucleon-nucleon interaction
of the form
$$
W({\bf r}_a - {\bf r}_b) = -\frac{g_s}{8\pi m_p}\left[ (g_0\mbox{\boldmath
 $\tau$}_a\cdot\mbox{\boldmath $\tau$}_b   +
g_2(\mbox{\boldmath $\tau$}_a\cdot \mbox{\boldmath $\tau$}_b
 -3\tau_a^z\tau_b^z))   ( \mbox{\boldmath $\sigma$}_a -
\mbox{\boldmath $\sigma$}_b)\right.
$$
\begin{equation} \label{4}
\left. + g_1(\tau_a^z\mbox{\boldmath $\sigma$}_a  - \tau_b^z\mbox{\boldmath
$\sigma$}_b)\right]  \cdot \mbox{\boldmath $\nabla$}_a \frac{e^{-m_{\pi}
r_{ab}}}{r_{ab}},
\end{equation}
where $m_p$ is the proton mass and $r_{ab}=|{\bf r}_a-{\bf r}_b|$. 

Together with the finite range interaction Eq.(\ref{4}) a
phenomenological zero range effective interaction has often been used in
estimates of the Schiff moment. It has the form \cite{khr91} 
$$
W_c( {\bf r}_a - {\bf r}_b) = \frac{G}{\sqrt{2}} \frac{1}{2m_p} \left(
(\eta_{ab}\mbox{\boldmath $\sigma$}_a - \eta_{ba} \mbox{\boldmath
$\sigma$}_b)\cdot \mbox{ \boldmath $\nabla$}_a \delta ({\bf r}_a - {\bf r}_b) +
\right.
$$
\begin{equation} \label{5}
\left. \eta_{ab}' [ \mbox{\boldmath $\sigma$}_a \times \mbox{\boldmath
$\sigma$}_b]\cdot\{({\bf p}_a - {\bf p}_b),\delta ({\bf r}_a - {\bf
r}_b)\}\right),
\end{equation}
where $G$ is the Fermi constant. The interaction (\ref{5}) can be
obtained from (\ref{4}) in the limit $m_\pi\rightarrow \infty$. In
this limit we have the following correspondence between the
interaction constants $\eta_{ab}$ and $g_i$ 
$$
\eta_{pp}=\frac{\sqrt{2}}{Gm_\pi^2}g_s(-g_0 +2g_2-g_1),\;\;
\eta_{nn}=\frac{\sqrt{2}}{Gm_\pi^2}g_s(-g_0+2g_2+g_1),
$$
\begin{equation} \label{6} 
\eta_{np}=\frac{\sqrt{2}}{Gm_\pi^2}g_s(g_0-2g_2+g_1),\;\;
\eta_{pn}=\frac{\sqrt{2}}{Gm_\pi^2}g_s(g_0-2g_2-g_1).
\end{equation}
The second term in Eq.(\ref{5}) comes from the exchange matrix
elements of the interaction (\ref{4}) in the limit $m_\pi \rightarrow \infty$ 
$$
\eta_{pp}'=\frac{\sqrt{2}}{Gm_\pi^2}g_s(g_0+2g_2+g_1),\;\;
\eta_{nn}'=\frac{\sqrt{2}}{Gm_\pi^2}g_s(g_0+2g_2-g_1),
$$
\begin{equation} \label{7}
\eta_{pn}'=\eta_{np}'=\frac{\sqrt{2}}{Gm_\pi^2}2g_s(g_0-g_2).
\end{equation}
One should however remember that the correspondence given by (\ref{6})
and (\ref{7}) is exact only in the limit of infinite pion mass. For
finite pion mass this correspondence is somewhat conditional since the
matrix elements of finite range interaction differ from the matrix
elements of zero range interaction. Especially large is the difference 
between the exchange matrix elements.

For a nucleus having an odd proton there are two kinds of
contributions to the nuclear Schiff moment even in the absence of a
strong residual interaction between the odd proton and the nucleons in
the core \cite{sfk}. One contribution comes from the odd proton.  We call it a single particle contribution.
Another contribution comes from the core nucleons.

The weak interaction, Eq.(\ref{4}), generates a weak correction to the nuclear
mean field. The correction is \cite{ds1}
$$
 \delta U({\bf r}) = \frac{g_sm_\pi^2}{\pi m_p} (\mbox{\boldmath
$\sigma$}\cdot {\bf n})\tau^z \int_0^\infty r'^2dr'\;
b(r,r')[(g_0-2g_2)(\rho_p(r)-\rho_n(r)) +
$$
\begin{equation} \label{8}
 g_1( \rho_p(r)+\rho_n(r))],
\end{equation}
where $b(r,r')$ is a combination of spherical Bessel functions of an
imaginary argument   $i_n(x)=\sqrt{\pi/2x} I_{n+1/2}(x)$, and $k_n(x)=\sqrt{2/\pi x} K_{n+1/2}(x)$ 
$$
b (r,r') = i_1(m_\pi r)k_0(m_\pi r') \theta(r'-r) -
i_0(m_\pi r')k_1(m_\pi r) \theta(r-r').
$$
Note, that the weak potential given by Eq.(\ref{8}) is proportional to $\tau^z$.
It is pure isovector. 
The contribution of $\Delta T=0,2$ channels to the weak potential is
proportional to $\rho_p-\rho_n$, while the contribution of $\Delta
T=1$  channel is proportional to $\rho_p+\rho_n$. It means that the
weak potential is most sensitive to $g_1$. The contributions of $g_0$
and $g_2$ are suppressed by the factor $(N-Z)/A$. This suppression has
an important effect on the $\Delta T=0$ channel. The finite range real
$NN$-interaction (\ref{4}) has both direct and exchange matrix
elements. The direct matrix elements contribute to the potential in
Eq.(\ref{8}). For a finite range interaction the exchange matrix
elements are usually small and in many cases can be safely
omitted. For the $\Delta T=0$ channel, due to suppression in the direct
potential, the exchange and direct potentials become comparable. 

The weak potential (\ref{8}) produces a correction $\delta\psi_\nu({\bf
r})$ to the single particle wave function $\psi_\nu({\bf r})$ of the
odd proton. With this correction the expectation value of the Schiff
moment is
\begin{equation} \label{10}
S_{sp}=\langle\psi_\nu +\delta\psi_\nu|\hat{S}_z|\psi_\nu +\delta\psi_\nu\rangle
= \langle\delta\psi_\nu|\hat{S}_z|\psi_\nu\rangle
+\langle\psi_\nu|\hat{S}_z|\delta \psi_\nu\rangle.  
\end{equation}
Note that this single particle contribution is absent for neutron odd
nuclei. In
Table 1 we show the single particle contributions to the Schiff moment
of $^{133}$Cs from the direct and exchange potentials in all three channels.  
\begin{table}
\caption{The contributions from the direct and exchange potentials to
the single particle Schiff moment of $^{133}$Cs. The units are $e$ fm$^3$.} 
\begin{center}
\begin{tabular}{||c|c|c|c||}
 &$g_s g_0$&$g_s g_1$&$g_s g_2$ \\
\hhline{----}
direct&$\;$ 0.011$\;$&$\;$ -0.109$\;$&$\;$ -0.022 \\
\hhline{----}
exchange &$\;$ -0.016$\;$ &$\;$ -0.003$\;$ & -0.006 \\ 
\end{tabular}
\end{center}
\end{table}
One can see that while in $\Delta T=1$ and 2 channels the exchange
contribution is really small, in $\Delta T=0$ channel it is even
larger in absolute value than the zero order term and has opposite sign. The same behaviour was
found also for $^{223}$Fr nucleus.  

The weak potential $\delta U$ is defined by diagonal matrix elements
of the two body weak interaction (\ref{4}) over core states. Another
type of contribution to the Schiff moment comes from off diagonal
matrix elements of the two body weak interaction. 
\begin{equation} \label{11}
S_{core}=\sum_{1p1h}\frac{\langle\nu,0|\hat{S}_z|1p1h,\nu\rangle\langle\nu,
1p1h|W|0,\nu\rangle +\langle\nu,0|W|1p1h,\nu\rangle\langle\nu,1p1h|
\hat{S}_z|0,\nu\rangle}{E_0-E_{1p1h}}.
\end{equation}
Calculating the matrix elements in (\ref{11}) we obtain the following
equation for this contribution
\begin{equation} \label{12}
S_{core}=\sum_{\nu_1\nu_2}\langle\nu,\nu_1|W|\nu_2,\nu\rangle
\frac{n_{\nu_1} - n_{\nu_2}}{\epsilon_{\nu_1} -
\epsilon_{\nu_2}}\langle\nu_2|\hat{S}_z |\nu_1\rangle,
\end{equation}
where $n_\nu$ are the occupation numbers and $\epsilon_\nu$ are the
single particle energies.The states $\nu_1$ and $\nu_2$ belong to
protons, therefore, for odd protons nuclei $S_{core}$ is proportional
to $\eta_{pp}$ while for odd neutrons nuclei it is proportional to
$\eta_{np}$. The magnitudes of $S_{sp}$ and $S_{core}$ are
comparable for proton odd nuclei. This is in contrast to the P odd and T
even  nuclear anapole moment where the core contribution is
 smaller by a factor $1/A^{1/3}$. In Table 2 we show the
singe particle and the core contributions for two proton odd nuclei.  
\begin{table}
\caption{The single particle and the core contributions for two proton
odd nuclei. Finite range weak interaction. The units are $e$ fm$^3$}
\begin{center}
\begin{tabular}{||c|c|c|c|c||}
& & $g_sg_0$&$g_sg_1$&$g_sg_2$ \\
\hhline{-----}
$^{133}$Cs& $S_{sp}$& -0.005&-0.112&-0.028 \\
\hhline{-----}
$^{133}$Cs&$S_{core}$ & 0.088&0.088& -0.176 \\
\hhline{-----}
$^{223}$Fr&$S_{sp}$&0.014& 0.084& 0.029 \\
\hhline{-----}
$^{223}$Fr&$S_{core}$&-0.135 & -0.135& 0.27 
\end{tabular}
\end{center}
\end{table}
The sum of three terms in the core contribution is proportional to the
combination $-g_0-g_1+2g_2 \sim \eta_{pp}=-\eta_{np}$.

At this place it is worth to compare the single-particle and the core contributions calculated in the limit of  zero range weak interaction.  The results of calculations are presented in Table III.
\begin{table}
\caption{The single particle and the core contributions for two proton
odd nuclei. Zero range weak interaction. The units are $e$ fm$^3$}
\begin{center}
\begin{tabular}{||c|c|c|c|c||}
& & $g_sg_0$&$g_sg_1$&$g_sg_2$ \\
\hhline{-----}
$^{133}$Cs& $S_{sp}$& -0.06&-0.23&-0.0562 \\
\hhline{-----}
$^{133}$Cs&$S_{core}$ & 0.18&0.18& -0.36 \\
\hhline{-----}
$^{223}$Fr&$S_{sp}$&0.071& 0.19& 0.072\\
\hhline{-----}
$^{223}$Fr&$S_{core}$&-0.209 & -0.209& 0.418 
\end{tabular}
\end{center}
\end{table}
We see that the difference is approximatly by a factor 2. The matrix elements of a finite range interaction are smaller due to smaller overlap of the wave functions in matrix elements of the interaction. However, this is not a general rule.  For the Hg nucleus, as we mentioned in \cite{ds1} the effect of a finite range is insignificant and goes even in opposite direction. For a finite range interaction we obtained
$$
S_{core} =-0.085 (g_sg_0 + g_sg_1 -2g_sg_2),
$$
while for zero range interaction
$$
S_{core} =-0.058 (g_sg_0 + g_sg_1 -2g_sg_2).
$$
For Rn nucleus the difference is 38\% and for Xe nucleus it becomes again close to 2.

\section{Core polarization}
The above results were obtained in a simple independent particle
model (IPM). In our calculations we used a full single-particle spectrum including continuum. 
The single particle basis was obtained using the partially self-consistent mean-field potential of \cite{BS74}. The potential includes four terms. The isoscalar term is the standard Woods-Saxon potential
\begin{equation} \label{12'}
U_0(r)=-\frac{V}{1+exp[(r-R)/a]},
\end{equation}
with the parameters $V=52.03$ MeV, $R=1.2709A^{1/3}$ fm, and $a=0.742$ fm. Two other terms $U_{ls}(r)$ and $U_\tau (r)$ were obtained in a self-consistent way using the two-body Migdal-type interaction \cite{mig} for the spin-orbit and isovector part of the potential. The last term is the Coulomb potential of a uniformly charged sphere with $R_c=1.18A^{1/3}$ fm. 

In the next step, the residual strong interaction between the odd
particle and the particles in the core should be taken into account.
In IPM the state of an odd nucleus can be presented as 
$$
|\nu\rangle = a^\dagger_\nu|0\rangle,
$$
where $|0\rangle$ is the ground state of an even core. When the
interaction is switched on the odd nucleus state becomes a
complicated superposition of the exited states of the core and the odd
particle 
\begin{equation} \label{13}
\left.|\nu\right) = A a^\dagger_\nu|0\rangle+B_{\nu'\nu}(1p1h)a^\dagger_{\nu'} 
|1p1h\rangle + C_{\nu'\nu}(2p2h)a^\dagger_{\nu'}|2p2h\rangle + \cdots.
\end{equation}
(Note that the round brackets mean a perturbed state).
This superposition can be conveniently written as
\begin{equation} \label{14}
\left.|\nu\right)=\hat{U}a^\dagger_\nu|0\rangle,
\end{equation}
where $\hat{U}$ is a unitary transformation depending on a nucleon-nucleon interaction. The matrix elements of any operator $\hat{O}$
between the exact states of the odd nucleus can
be expressed by matrix elements of an effective operator
$\hat{\tilde{O}}$ between the IPM states
\begin{equation} \label{15}
(\nu'|\hat{O}|\nu) = \langle\nu'|\hat{\tilde{O}}|\nu\rangle,
\end{equation}
where $\hat{\tilde{O}}=U^\dagger\hat{O}U$ is the effective operator.
For a one body operator, like the Schiff moment, in first order in the nucleon-nucleon interaction the matrix elements of the effective operator coincide
 with Eq.(\ref{12}) for the core contribution.  
\begin{equation} \label{16}
\langle\nu'|\tilde{O}|\nu\rangle =\langle\nu'|O|\nu\rangle+\sum_{\nu_1\nu_2}
\langle\nu'\nu_1|V|\nu_2\nu\rangle
\frac{n_{\nu_1} - n_{\nu_2}}{\epsilon_{\nu_1} -
\epsilon_{\nu_2}}\langle\nu_2|O|\nu_1\rangle,
\end{equation}
where $V$ is the full nucleon-nucleon residual interaction. Leaving only
1p1h intermediate states we can sum over perturbation series obtaining
for the effective Schiff moment the equation 
\begin{equation} \label{17}
\langle\nu'|\tilde{\bf S}|\nu\rangle =\langle\nu'|{\bf S}|\nu\rangle +\sum_{\nu_1\nu_2}
\langle\nu'\nu_1|V|\nu_2\nu\rangle
\frac{n_{\nu_1} - n_{\nu_2}}{\epsilon_{\nu_1} -
\epsilon_{\nu_2}}\langle\nu_2|\tilde{\bf S}|\nu_1\rangle,
\end{equation}
where $\langle\nu'|{\bf S}|\nu\rangle$ is the single particle matrix element of the
bare Schiff moment operator, Eq.(\ref{00}). Eq.(\ref{17}) is just the RPA
equation for the effective field in the terminology of Migdal's theory
\cite{mig}. This equation describes the effects of core polarization
due to interaction of the odd particle with the particles in the core.
 
Diagonal matrix elements of the Schiff moment are nonzero only in the
presence of P and T violating interactions. For this reason the full
nucleon nucleon interaction $V$ must include this interaction 
$$
V({\bf r}_1,{\bf r}_2)= F({\bf r}_1,{\bf r}_2)+W({\bf r}_1-{\bf r}_2),
$$
where $F$ is the strong residual interaction and $W$ is given by
Eq.(\ref{4}). In the same way, the mean field potential is the sum of
two terms $U(r)+\delta U({\bf r})$, where  $U(r)$ is the main mean
field created by the strong interaction and $\delta U({\bf r})$ is the
weak correction given by Eq.(\ref{10}). As  metioned
above, $\delta U({\bf r})$ creates the correction $\delta\psi_\nu({\bf
r})$ for every single particle wave function $\psi_\nu({\bf r})$ in
all matrix elements in Eq.(\ref{17}). Since the corrections are small
we can retain only the first order. To do this one should perform the
substitutions $V \rightarrow F+W$ and all $\psi_\nu \rightarrow
\psi_\nu + \delta \psi_\nu$ in Eq.(\ref{17}) and keep the terms linear
in $W$ or $\delta \psi$. After this procedure the contributions to
the Schiff moment can be written as a sum of three terms
\begin{equation} \label{18}
S = \langle\delta\psi_\nu|\tilde{S}_z|\psi_\nu\rangle +
\langle\psi_\nu|\tilde{S}_z|\delta\psi_\nu\rangle +\langle\nu|\delta S|\nu\rangle. 
\end{equation}
The first two terms are those where the weak correction enters via the
odd particle wave function. $\tilde{S}$ satisfies the equation
\begin{equation} \label{19}
\langle\nu'|\tilde{\bf S}|\nu\rangle =\langle\nu'|{\bf S}|\nu\rangle+\sum_{\nu_1\nu_2}
\langle\nu'\nu_1|F|\nu_2\nu\rangle
\frac{n_{\nu_1} - n_{\nu_2}}{\epsilon_{\nu_1} -
\epsilon_{\nu_2}}\langle\nu_2|\tilde{\bf S}|\nu_1\rangle,
\end{equation}
that differs from Eq.(\ref{17}) only in the interaction. Here only
the residual strong interaction $F$ enters in the equation. The effects of
the weak interaction are entirely in the wave functions of the odd
proton. Eq.(\ref{19}) describes the well known effect of renormalization
of nuclear moments due to coupling with particle-hole states in the
core. Renormalization of the Schiff moment due to coupling with the
isoscalar dipole modes were discussed in \cite{hb00}. It was found there that
the effect of renormalization for $^{209}$Bi nucleus is not
significant, about 15\%. We found similar values both for Cs and Fr
nuclei for renormalization of the  isoscalar component of the Schiff moment.  The isovector componet, however, undergoes significant renormalization. Due to strong repulsion in the isovector channel it leads to reduction of the isovector component.  As a result, the single particle contrubution becomes reduced by the factor $\sim 2$. For Cs, instead of values cited in the first line in Table I we obtain for the renormalized single particle contribution $\tilde{S}_{sp}$
$$
\tilde{S}_{sp} = -0.003g_sg_0 - 0.056g_sg_1 - 0.016g_sg_2.
$$

The third term in (\ref{18}) satisfies the equation  
\begin{equation} \label{20}
\langle\nu'|\delta {\bf S}|\nu\rangle =\langle\nu'|\delta {\bf
S}_0|\nu\rangle+\sum_{\nu_1\nu_2} 
\langle\nu'\nu_1|F|\nu_2\nu\rangle
\frac{n_{\nu_1} - n_{\nu_2}}{\epsilon_{\nu_1} -
\epsilon_{\nu_2}}\langle\nu_2|\delta{\bf S}|\nu_1\rangle,
\end{equation}
that looks similar to (\ref{19}). However, the inhomogenious term
$\delta {\bf S}_0$ is completely different, namely
$$
\langle\nu'|\delta{\bf S}_0|\nu\rangle=\sum_{\nu_1\nu_2} 
\langle\nu'\nu_1|W|\nu_2\nu\rangle
\frac{n_{\nu_1} - n_{\nu_2}}{\epsilon_{\nu_1} -
\epsilon_{\nu_2}}\langle\nu_2|\tilde{\bf S}|\nu_1\rangle  +
$$
$$
\sum_{\nu_1\nu_2} 
(\langle\nu'\delta\psi_{\nu_1}|F|\nu_2\nu\rangle+
\langle\nu'\nu_1|F|\delta\psi_{\nu_2}\nu\rangle)  
\frac{n_{\nu_1} - n_{\nu_2}}{\epsilon_{\nu_1} -
\epsilon_{\nu_2}}\langle\nu_2|\tilde{\bf S}|\nu_1\rangle +
$$
\begin{equation} \label{21}
\sum_{\nu_1\nu_2} 
\langle\nu'\nu_1|F|\nu_2\nu\rangle
\frac{n_{\nu_1} - n_{\nu_2}}{\epsilon_{\nu_1} -
\epsilon_{\nu_2}}(\langle\delta\psi_{\nu_2}|\tilde{\bf S}|\nu_1\rangle
+ \langle\nu_2|\tilde{\bf S}|\delta\psi_{\nu_1}\rangle).
\end{equation}
The first term on the rhs of Eq.(\ref{21}) is the core contribution
(\ref{12}), where instead of the bare Schiff moment operator (\ref{00})
enters the renormalized operator $\tilde{\bf S}$. The second and the
third terms correspond to additional contributions where the weak
interaction enters via corrections to the intermediate single particle
states $|\nu_1>$ and $|\nu_2>$. Equations (20-22) describe all
 of the core polarization effects.  The main contribution to $\delta {\bf S}_0$ comes from the 
first term in Eq. (\ref{21}). This term depends on the renormalized Schiff moment $\tilde{\bf S}$, 
therefore its contribution becomes reduced proportionally to the reduction of $\tilde{\bf S}$, approximately by the factor $\sim 2$. The contribution of the second and the third terms in Eq.(\ref{21}) is small. It is noticable only in T=1 channel.

In Tables IV, and V we present the results of calculations of the
Schiff moment for sets of odd proton and odd neutron nuclei. The
effects of the core polarization are twofold. First, without core
polarization, for odd neutron nuclei the Schiff moment depended only
on one constant $\eta_{np}$ (see Eq.(\ref{6})). Now, the Schiff moment depends on all
three constants. Second, the values of the Schiff moment are
decreased. There are two sources of that reduction. First, it is mentioned above reduction in
the first term in Eq.(\ref{21}).  Further decrease comes from repulsive nature of spin-spin
residual forces responsible for renormalization of $\delta{\bf S}_0$
in Eq.(\ref{20}).     
\begin{table}
\caption{Effects of core polarization on the Schiff moment for odd
proton nuclei, for finite range weak interaction. The bare values of the Schiff moment, without core
polarization, are listed in the first line for each nucleus. The
units are $e\, fm^3$}
\begin{center}
\begin{tabular}{||c|c|c|c||}
&$g_sg_0$& $g_sg_1$& $g_sg_2$ \\
\hhline{----}
$^{133}$Cs$\;$ & $\;\;$0.08$\;$&$\;\;$-0.02$\;$&$ \;\;\;$ -0.21$\;\;\;$ \\
\hhline{----}
$^{133}$Cs$\;$ & $\;\;$0.006$\;$&$\;\;$-0.02$\;$&$ \;\;\;$ -0.04$\;\;\;$ \\
\hhline{----}
$^{223}$Fr$\;$ & -0.122 & -0.052 & 0.300 \\
\hhline{----} $^{223}$Fr$\;$ & -0.009 & -0.016 & 0.030
\end{tabular}
\end{center}
\end{table}

\begin{table}
\caption{Effects of core polarization on the Schiff moment for
neutron odd nuclei, for finte range weak interaction. The bare values of the Schiff moment, without
core polarization, are listed in the first line for each nucleus. The units are
$e\, \mbox{fm}^3$.}
\begin{center}
\begin{tabular}{||c|c|c|c||}
 & $g_sg_0$ & $g_sg_1$& $g_sg_2$ \\
\hhline{----}
$^{199}$Hg$\;$&$\;\; -0.09 \;$&$\;$ -0.09$\;$&$\;\;$ 0.18$\;$\\ 
\hhline{----}
$^{199}$Hg$\;$&$\;$ -0.00004$\;$ &$\;$ -0.055$\;$&$\;\;$ 0.009$\;$ \\
\hhline{----}
$^{129}$Xe$\;$&$\;$ 0.06$\;$ &$\;$0.06$\;$&$\;\;$-0.12$\;$\\
\hhline{----}
$^{129}$Xe$\;$&$\;$ 0.008$\;$ &$\;$0.006$\;$&$\;\;$-0.009$\;$\\
\hhline{----} 
$^{211}$Rn$\;$&$\;$ -0.12$\;$&$\;$-0.12$\;$&$\;\;$0.24$\;$\\
\hhline{----} 
$^{211}$Rn$\;$&$\;$ -0.019$\;$&$\;$0.061$\;$&$\;\;$0.053$\;$\\
\hhline{----} 
$^{213}$Ra$\;$&$\;$ -0.12$\;$&$\;$-0.12$\;$&$\;\;$0.24$\;$\\
\hhline{----} 
$^{213}$Ra$\;$&$\;$ -0.012$\;$&$\;$-0.021$\;$&$\;\;$0.016$\;$\\
\hhline{----}
$^{225}$Ra$\;$&$\;$ 0.08$\;$ &$\;$0.08$\;$&$\;\;$-0.16$\;$\\
\hhline{----}
$^{225}$Ra$\;$&$\;$ 0.033$\;$ &$\;$-0.037$\;$&$\;\;$-0.046$\;$\\
\end{tabular}
\end{center}
\end{table}

\section{Conclusions}
We stress here two main conclusions from this work. First, the core-polarization contributions to the Schiff moments are sizeable. These corrections reduce the lowest order values obtained in the past by factors two to three. This has some important  consequences when trying to interpret the implication of the experimental atomic dipole measurements. The reductions of the Schiff moments (compared to the values obtained in the past) means that using the experimental  upper limits  for the atomic dipole one obtains now a higher upper limit for the strength parameters in  the weak
parity and time reversal  non-conserving component of the N-N interaction.
Second, we have seen that the finite range reduces the mixing matrix elements 
between  the opposite parity states in the case of spherical nuclei. 
It was argued \cite{eng03} that the large enhancement factors for the Schiff moments found in quadrupole+octupole nuclei \cite{spe97}  will be reduced when finite range interactions are used. We see that the enhancement factors (compared to the spherical case) will not be reduced because the effect of finite range has a similar effect in spherical and  in deformed nuclei.

We wish to thank V. Zelevinsky for helpful discussions. One of us (V.D.) wants to thank Tel-Aviv University for the hospitality. This work was supported by the US -- Israel Binational Science Foundation.

\end{document}